\documentclass[%
reprint,
superscriptaddress,
amsmath,amssymb,
prc,
floatfix, ]%
{revtex4-1}
\usepackage{color}
\usepackage{graphicx}
\usepackage{dcolumn}
\usepackage{bm}

\newcommand{\element}[2]{~${}^{#1}${#2}}

\begin{document}


\title{The alternating-parity bands of \element{236,238}{U} and \element{238,240}{Pu} in a particle-number conserving method based on cranked shell model}

\author{Xiao-Tao He (贺晓涛)}
\email{hext@nuaa.edu.cn}
 \affiliation{College of Materials Science and Technology, Nanjing University of
              Aeronautics and Astronautics, Nanjing 210016, China}
\author{Yu-Chun Li (李玉春)}
 \affiliation{Strategic Technology Equipment Development Center, China Academy of Engineering Physics, Mianyang 621900, China}

\date{\today}

\begin{abstract}
The particle-number conserving (PNC) method in the framework of cranked shell model (CSM) is developed to deal with the reflection-asymmetric nuclear system by applying the $S_x$ symmetry. Based on an octupole-deformed Nilsson potential, the alternating-parity bands in \element{236,238}{U} and \element{238,240}{Pu} are investigated. The experimental kinematic moments of inertia (MoI) and the angular momentum alignments of all studied bands are reproduced well in the PNC-CSM calculations. The striking difference of rotational behaviors between U and Pu isotopes can be linked to the strength of octupole correlations. The upbendings of the alternating-parity bands in\element{236,238}{U} are due to the alignments of pairs of nucleons occupying $\nu g_{9/2}$, $\pi f_{7/2}$ orbitals and $\nu j_{15/2}$,  $\pi i_{13/2}$ high-$j$ intruder orbitals. Particularly, the interference terms of nucleon occupying the octupole-correlation pairs of $\nu^2 j_{15/2} g_{9/2}$ and of $\pi^2 i_{13/2} f_{7/2}$ give a very important contribution to the suddenly gained alignments.  

\end{abstract}

\maketitle

\section{Introduction}

Since the first observation of low lying negative parity states by the Berkeley group in the 1950s~\cite{AsaroF1953_PR92_1495, StephensF1954_PR96_1568}, octupole correlations have long been an attractive topic in nuclear structure physics~\cite{ChasmanR1980_PLB96_7, AhmadI1993_ARNPS43_71, ButlerP1996_RMP68_349,GaffneyL2013_N497_199}.  In a microscopic picture, these result from the long range octupole interaction between normal orbital with quantum numbers ($l-3, j-3$) and intruder orbital of opposite parity with quantum numbers ($l, j$). When these pairs occupy states near the Fermi surface, strong octupole correlations can lead to reflection asymmetric shapes. Nuclei with $Z\approx 34, 56, 88$ and $N\approx 34, 56, 88, 134$ are expected to meet the condition~\cite{ButlerP1996_RMP68_349}. Many experimental phenomena provided indications for reflection asymmetric deformation, such as alternating-parity bands in even-even nuclei~\cite{SmithJ1995_PRL75_1050, CocksJ1997_PRL78_2920}, parity doublets in odd-A nuclei~\cite{DahlingerM1988_NPA484_337}, and connected enhanced electric dipole transitions inter these bands.

Several theoretical approaches have been developed to study the properties of reflection asymmetric shape and rotational bands (see Ref.~\cite{ButlerP1996_RMP68_349} and reference therein). These include the macroscopic-microscopic models~\cite{NazarewiczW1984_NPA429_269, MoellerP2008_ADNDT94_758}, self-consistent mean field models~\cite{BoncheP1986_PLB175_387, EgidoJ1991_NPA524_65, BenderM2004_PRC70_54304, RobledoL2011_PRC84_54302, RobledoL2013_PRC88_51302}, cranking model~\cite{NazarewiczW1985_NPA441_420}, interacting boson models~\cite{EngelJ1987_NPA472_61, KusnezovD1988_PLB209_420, CottleP1998_PRC58_1500, ZamfirN2003_PRC67_14305}, cluster models~\cite{BuckB1996_PRL76_380, ShneidmanT2002_PLB526_322, ShneidmanT2003_PRC67_14313, ShneidmanT2006_PRC74_34316, ShneidmanT2015_PRC92_34302},
phenomenological collective models~\cite{BizzetiP2004_PRC70_64319, LenisD2006_PLB633_474, BizzetiP2010_PRC81_34320, MinkovN2012_PRC85_34306, BonatsosD2015_PRC91_54315}
and reflection asymmetric shell model~\cite{ChenY2000_PRC63_14314, ChenY2008_PRC77_61305} etc. Cranked shell model is one of the most useful microscopic model to investigate the nuclear rotational collectivity. Specifically, because simplex operator commutes with cranking Hamiltonian, it is very convenient to study the properties of rotational bands with octupole correlations. 

Pairing correlations are essential to describe not only the nuclear ground-state properties but also the excited state properties~\cite{BohrA1975_2,RingP1980}. In the framework of cranked shell model, a particle-number conserving (PNC-CSM) method is used to treat the pairing correlations~\cite{ZengJ1983_NPA405_1,WuC1989_PRC39_666,ZengJ1994_PRC50_1388,XinX2000_PRC62_67303,HeX2018_PRC98_64314}. In the PNC-CSM method, the cranked shell model Hamiltonian is diagonalized directly in a truncated Fock space and a pair-broken excited configuration is defined by blocking the real particles in the single-particle orbitals~\cite{WuC1989_PRC39_666}. The particle number is conserved and the Pauli blocking effect is treated spontaneously. The PNC-CSM method has previously been applied successfully to describe the intrinsic reflection-symmetric system from the light nuclear mass region around $A=40$~\cite{XiangX2018_CPC42_54105} to the very heavy region around $A=250$~\cite{ZhangZ2012_PRC85_14324,LiY2016_SCPMA59_672011,HeX2020_CPC44_34106,HeX2005_NPA760_263}. 

Previous PNC-CSM method can not be applied to study the reflection-asymmetric nuclear system. In the present work, the PNC-CSM method is developed to include the octupole deformation and then used to investigate the alternating-parity rotational bands in \element{236,238}{U} and \element{238,240}{Pu}. Actinide region is one of the typical nuclear mass region where signatures of octupole correlations have been identified in the experiment~\cite{AhmadI1993_ARNPS43_71,ButlerP1996_RMP68_349}. It is found experimentally that the rotational behaviors between \element{236,238}{U} and \element{238,240}{Pu} are dramatically different and this issue needs further theoretical investigations~\cite{WardD1996_NPA600_88,WiedenhoeverI1999_PRL83_2143,WangX2009_PRL102_122501,ZhuS2010_PRC81_41306,SpiekerM2013_PRC88_41303}. In the present work, the striking difference of the rotational properties in \element{236,238}{U} and \element{238,240}{Pu} is explained.

The PNC-CSM formalism for reflection-asymmetric shapes by applying the $S_x$ symmetry are presented in Sec.~\ref{sec:PNC_CSM}. In Sec.~\ref{sec:Exp}, arguments based on experimental alternating-parity bands are presented for the stable octupole deformation at high spins in U and Pu isotopes. The results of the PNC-CSM calculations based on an octupole-deformed Nilsson potential and the discussions of the microscopic mechanism which leads to the different rotational behaviors between U and Pu isotopes are presented in Sec.~\ref{sec:results}. Finally, a brief summary are given in Sec.~\ref{sec:summary}

\section{PNC-CSM formalism in the presence of octupole deformation}{\label{sec:PNC_CSM}}
\subsection{Cranked single-particle levels}
The CSM Hamiltonian of an axially deformed nucleus in the rotating frame is,
\begin{eqnarray}
 H_\mathrm{CSM}
 & = &
 H_0 + H_\mathrm{P}
 = H_{\rm Nil}-\omega J_x + H_\mathrm{P}\ ,
 \label{eq:H_CSM}
\end{eqnarray}
where $-\omega J_x=-\omega\sum j_x$ is the Coriolis interaction with the cranking frequency $\omega$ about the $x$ axis. Note that rotations about different axes ($x$ or $y$ axes) perpendicular to the symmetry axis ($z$ axis here) are equivalent. For definiteness, we choose rotation about the $x$ axis to discuss. $H_{\rm Nil} = \sum h_{\rm Nil} $ is the Nilsson Hamiltonian~\cite{NilssonS1969_NPA131_1,JohanssonS1961_NP22_529,VogelP1967_PLB25_65},
\begin{multline}
h_{\rm Nil}=\frac{1}{2}\hbar\omega_0(\varepsilon_2,\varepsilon_3,\varepsilon_4)
        \left[-\nabla^2_\rho
          +\frac{1}{3}\varepsilon_2
                                  \left(2\frac{\partial^2}{\partial\zeta^2}-
                                   \frac{\partial^2}{\partial\xi^2}-
                                   \frac{\partial^2}{\partial\eta^2}\right) \right. \\
     \left.  +\rho^2
          -\frac{2}{3}\varepsilon_2\rho^2P_2(\cos\theta_t)
          +2\varepsilon_3\rho^2P_3(\cos\theta_t)
          +2\varepsilon_4\rho^2P_4(\cos\theta_t)
           \right]\\
         -2\kappa\hbar\mathring{\omega}_0
         \left[\vec{l_t}\cdot\vec{s}-
             \mu\left(\vec{l_t}^2-\langle\vec{l_t}^2\rangle_N\right)
         \right ],
\end{multline}
where $\varepsilon_2, \varepsilon_3, \varepsilon_4$ are the quadrupole, octupole and hexadecapole deformation parameters, respectively, and the subscript $t$ means that the single particle Hamiltonian $h_{\rm Nil}$ is written in the stretched coordinates ($\xi,\eta,\zeta$). 

In the spherical harmonic oscillation basis $|Nl\Lambda\Sigma\rangle$, they are the corresponding quantum numbers of $N,l^2,l_z,s_z$, respectively, the selection rules of the matrix elements of $\rho^2P_3(\cos\theta_t)$ are,
\begin{eqnarray}
N'=N\pm1,\ \ \ l'=l\pm3,\ \ \ \Lambda'=\Lambda,\ \ \ \Sigma'=\Sigma\ .
\end{eqnarray}
Since parity $p=(-1)^N$, both symmetries of space inversion $P$ and rotation $R_{x}(\pi)=e^{-i\pi j_x}$ are broken in an intrinsic reflection asymmetric system ($\varepsilon_3\neq0$). However, in this case, the reflection through planes $yoz$, $S_{x}$ invariant holds~\cite{BohrA1975_2}. According to the definition of Goodman, $S_{x}= PR_{x}^{-1}(\pi)$~\cite{GoodmanA1974_NPA230_466}. 

Due to the Coriolis interaction $-\omega j_x$, $\Omega=\Lambda+\Sigma$ (single-particle angular momentum projection on the symmetry axis $z$) is a good quantum number no longer. However, $[j_x,j^2_z]=0$ and $[S_x, j_{z}^2] = 0$ hold still. The eigenstates of $h(\omega)=h_{\mathrm{Nil}}-\omega j_x$ can be characterized by the simplex $s$ (the eigenvalue of $S_x$). A good-simplex basis can be constructed in a reflection asymmetric system. Let $|\xi\rangle$ denotes the spherical single-particle basis $|N_{\xi}l_{\xi}\Lambda_{\xi}\Sigma_{\xi}\rangle$, and $|\bar{\xi}\rangle=T|\xi\rangle$ its time-reversal state. A new single-particle basis is obtained by transform the time-reversal basis $|\xi\rangle$ ($|\bar{\xi}\rangle$) to simplex basis $|\xi s\rangle$, 
\begin{eqnarray}
\nonumber
|\xi,s=\pm i\rangle&=&\frac{1}{\sqrt2}\left[|\xi\rangle\mp iS_x|\xi\rangle\right]\\
\nonumber
&=&\frac{1}{\sqrt2}\left[|\xi\rangle\pm (-1)^{\Omega_\xi-1/2}T|\xi\rangle\right]\\
&=&\frac{1}{\sqrt2}\left[a^\dag_\xi\pm(-1)^{\Omega_\xi-1/2}a^{\dag}_{\bar{\xi}}\right]|0\rangle
\end{eqnarray}
where $|\xi\rangle=a^\dag_\xi|0\rangle$, $|\bar{\xi}\rangle=a^\dag_{\bar{\xi}}|0\rangle$. The creation operator for a nucleon on state $|\xi s\rangle$ is $\beta^\dag_{\xi s}=\frac{1}{\sqrt2}\left[a^\dag_\xi\pm(-1)^{\Omega_\xi-1/2}a^{\dag}_{\bar{\xi}}\right]$. $|\xi s\rangle$ is the simultaneous eigenstate of $S_{x}$ and $j^2_z$,
\begin{eqnarray}
S_{x}|\xi s\rangle&=&s|\xi s\rangle\ ,\ \ \ s=\pm i,\\
j^2_z|\xi s\rangle&=&\Omega^2_\xi|\xi s\rangle.
\end{eqnarray}
The non-zero matrix elements of $h(\omega)=h_{\mathrm{Nil}}-\omega j_x$ are
\begin{eqnarray}
\langle\xi s|h_{\mathrm{Nil}}|\xi' s'\rangle=\langle\xi |h_{\mathrm{Nil}}|\xi'\rangle\delta_{ss'},
\end{eqnarray}
and,
\begin{equation}
\langle\xi s|j_x|\xi' s'\rangle=\Bigg\{
\begin{aligned}
 &\langle\xi |j_x|\xi'\rangle\delta_{ss'},\ \ \ \Omega_{\xi}\neq\frac{1}{2}\ \mathrm{or}\ \Omega_{\xi'}\neq\frac{1}{2},\\       
 &\pm\langle\xi |j_x|-\xi'\rangle\delta_{ss'},\ \ \ \Omega_{\xi}=\Omega_{\xi'}=\frac{1}{2}\ .
 \end{aligned}
\end{equation}
By diagonalizing $h(\omega)$ in the $|\xi s\rangle$ basis, the eigenstates $|\mu s\rangle$ of the cranked single-particle Hamiltonian is expressed as
\begin{equation}
|\mu s\rangle=\sum_\xi C_{\mu\xi}(s)|\xi s\rangle,\ \ \ C_{\mu\xi}(s)\ \textrm{is real} 
\end{equation}
Hereafter, $|\mu s\rangle$ is sometimes denoted simply by $|\mu\rangle$. $b^\dag_{\mu s}=\sum_\xi C_{\mu\xi}(s)\beta^\dag_{\xi s}$ is the \textit{real} particle operator of the cranked single-particle state $|\mu s\rangle$. 

\subsection{Cranked many-particle configuration}

For a n-particle system, the cranked many-particle configuration (CMPC) is
\begin{eqnarray}
|i\rangle=|\mu_1\mu_2\cdots\mu_n\rangle=b^\dag_{\mu_1}b^\dag_{\mu_2}\cdots b^\dag_{\mu_n}|0\rangle,
\end{eqnarray}
$\mu_1\mu_2\cdots\mu_n$ are the occupied cranked Nilsson orbitals. 
Each configuration is characterized by the simplex $s$,
\begin{eqnarray}
s_i=\prod_{\mu_i(occ.)}s_{\mu_{i}},
\end{eqnarray}
and the energy of each configuration is,  
\begin{eqnarray}
E_i=\sum_{\mu_i(occ.)}\epsilon_{\mu_{i}}.
\end{eqnarray}
where $\mu_i(occ.)$ denotes the occupied cranked Nilsson orbitals. 

\subsection{Pairing correlations}
The pairing includes the monopole- and quadrupole-pairing correlations $H_\mathrm{P}(0)$ and $H_\mathrm{P}(2)$,
\begin{eqnarray}
 H_{\rm P}(0)
 & = &
  -G_{0} \sum_{\xi\eta} a^\dag_{\xi} a^\dag_{\bar{\xi}}
                        a_{\bar{\eta}} a_{\eta} \ , \\
 H_{\rm P}(2)
 & = &
  -G_{2} \sum_{\xi\eta} q_{2}(\xi)q_{2}(\eta)
                        a^\dag_{\xi} a^\dag_{\bar{\xi}}
                        a_{\bar{\eta}} a_{\eta}\ ,
\end{eqnarray}
where $q_{2}(\xi) = \sqrt{{16\pi}/{5}}\langle \xi |r^{2}Y_{20} | \xi \rangle$ is the diagonal element of
the stretched quadrupole operator, and $G_0$ and $G_2$ are the effective strengths of monopole- and quadrupole-pairing interactions, respectively.

In the cranked simplex representation, 
\begin{eqnarray}
H_P(0)&=&-G_0\sum_{\mu\mu{\prime}\nu\nu{\prime}}f^{\ast}_{\mu\mu{\prime}}f_{\nu{\prime}\nu}
b_{\mu+}^{\dagger}b_{\mu{\prime}-}^{\dagger}b_{\nu-}b_{\nu{\prime}+},\\
\nonumber f^{\ast}_{\mu\mu{\prime}}&=&\sum_{\xi>0}(-)^{\Omega_\xi}C_{\mu\xi}(+)C_{\mu{\prime}\xi}(-), \\ 
\nonumber f_{\nu{\prime}\nu}&=&\sum_{\eta>0}(-)^{\Omega_\eta}C_{\nu{\prime}\eta}(+)C_{\nu\eta}(-).
\end{eqnarray}
and
\begin{eqnarray}
H_P(2)&=&-G_2\sum_{\mu\mu{\prime}\nu\nu{\prime}}g^{\ast}_{\mu\mu{\prime}}g_{\nu{\prime}\nu}
b_{\mu+}^{\dagger}b_{\mu{\prime}-}^{\dagger}b_{\nu-}b_{\nu{\prime}+},\\
\nonumber g^{\ast}_{\mu\mu{\prime}}&=&\sum_{\xi>0}(-)^{\Omega_\xi}C_{\mu\xi}(+)C_{\mu{\prime}\xi}(-)q_2(\xi), \\
\nonumber g_{\nu{\prime}\nu}&=&\sum_{\eta>0}(-)^{\Omega_\eta}C_{\nu{\prime}\eta}(+)C_{\nu\eta}(-)q_2(\eta).
\end{eqnarray}
where $C_{\mu\xi}(+)$ ($b_{\mu+}^{\dagger}$) and $C_{\mu\xi}(-)$ ($b_{\mu-}^{\dagger}$) stand for the case of $s=+i$ and $s=-i$, respectively.  

\subsection{Particle-number conserving method}

The cranked shell model Hamiltonian $H_{\rm{CSM}}$ is diagonalized in a sufficiently large cranked
many-particle configuration space and then sufficiently accurate low-lying excited eigenstates are obtained as,
\begin{equation}
 | \psi \rangle = \sum_{i} C_i | i \rangle,
 \label{eq:eigenstate}
\end{equation}
where $| i \rangle$ is a cranked many-particle configuration of the n-body system and $C_i$ is the corresponding amplitude. Note that $|\psi\rangle$ is parity-independent but with certain simplex.

The occupation probability $n_{\mu}$ of each cranked Nilsson orbital $\mu$ can be calculated as,
\begin{equation}
\label{eq:n_mu}
    n_{\mu} = \sum_{i}|C_{i}|^{2}P_{i\mu},
\end{equation} 
where $P_{i\mu}=1$ if $|\mu\rangle$ is occupied and $P_{i\mu}=0$ otherwise. The total particle number $N=\sum_\mu n_\mu$. The configuration of a rotational band, including sidebands built on pair-broken excited intrinsic configurations, can be determined by the rotational frequency $\omega$-dependence of occupation probabilities $n_\mu$. 

\subsection{Moment of inertia}
The angular momentum alignment includes the diagonal and off-diagonal parts,
\begin{equation}
 \langle \psi | J_x | \psi \rangle = \sum_i |C_i|^2 \langle i | J_x | i \rangle + 2\sum_{i<j}C_i^{*} C_j \langle i | J_x | j \rangle \ .
 \label{eq:jx}
\end{equation}
$ \langle  \psi | J_x | \psi \rangle$ is simplified as $\langle J_x \rangle$ hereafter sometimes. $J_x$ is an one-body operator, the off-diagonal parts $\langle i | J_x |j \rangle$ ($i \neq j$) does not vanish only when $|i\rangle$ and $|j\rangle$ differ by one particle occupation. After a certain permutation of creation operators, $|i\rangle$ and $|j\rangle$ are reconstructed into,
\begin{equation}
 |i\rangle=(-1)^{M_{i\mu}}|\mu\cdots \rangle \ , \quad
 |j\rangle=(-1)^{M_{j\nu}}|\nu\cdots \rangle \ ,
\end{equation}
where the ellipsis stands for the same particle occupation, and $(-1)^{M_{i\mu}}=\pm1$, $(-1)^{M_{j\nu}}=\pm1$ depend on whether the permutation is even or odd. Then,
\begin{eqnarray}
\label{eq:Jx_single}
\nonumber
  && \langle J_x \rangle = \sum_{\mu}j_x(\mu)+\sum_{\mu<\nu}j_x(\mu\nu),\\
\nonumber
  && j_x(\mu) =  n_\mu\langle\mu|j_{x}|\mu\rangle,\\
\nonumber
  && j_x(\mu\nu) =  2\langle\mu|j_{x}|\nu\rangle\sum_{i<j} (-1)^{M_{i\mu}+M_{j\nu}} C_{i}^{*}C_{j}, (\mu\neq\nu),\\
\end{eqnarray}
where $j_x(\mu)$ is the diagonal contribution and $j_x(\mu\nu)$ the off-diagonal contribution.

The kinematic moment of inertia for the state $|\psi\rangle$ is given by
\begin{equation}
    J^{(1)} = \frac{1}{\omega} \langle \psi | J_x | \psi \rangle.
  \label{eq:moi}
\end{equation}

\subsection{Description for the octupole-defomred bands}\label{subsec:octband}
The square of the simplex operator $S_x$ is related to the total number of the nucleons,
\begin{equation}
S^2_x=(-1)^A
\end{equation}
The rotational band with simplex $s$ is characterized by spin states $I$ of alternation parity~\cite{BohrA1975_2},
\begin{equation}
    p = s \cdot e^{-i\pi I}.
\end{equation}
For reflection-asymmetric systems with even number of nucleons,
\begin{eqnarray}
s=+1,\  I^{p} = 0^+, 1^-, 2^+, 3^-, \cdots\ ,\\
s=-1,\  I^{p} = 0^-, 1^+, 2^-, 3^+, \cdots\ .
\end{eqnarray}
and for systems with odd number of nucleons,
\begin{eqnarray}
s=+i,\  I^p = 1/2^+, 3/2^-, 5/2^+, 7/2^-, \cdots ,\\
s=-i,\  I^p = 1/2^-, 3/2^+, 5/2^-, 7/2^+, \cdots .
\end{eqnarray}

In the limit of static octupole deformation, the properties of both rotational bands can have a unified description. The energies of the experimental alternating-parity bands in even-even nuclei and the parity doublet bands in odd-A or odd-odd nuclei can be described as~\cite{JolosR1995_NPA587_377},
\begin{equation}
    E(I) = E_{\rm av}(I) - \frac{1}{2} p \Delta E(I),
    \label{eq:E_exp}
\end{equation}   
$E_{\rm av}(I)$ is parity-independent energy of state $I$ in an intrinsic band,
\begin{equation}
    E_{\rm av}(I) = \frac{1}{2}[ E_{\rm inter}(I) + E_{\rm exp}(I) ],
\end{equation}
where $E_{\rm inter}(I)$ is a smooth interpolation between the energies of states in the positive parity band at the odd value of $I$ ~\cite{JolosR2015_PRC92_44318}, $E_{\rm exp}(I)$ is the energy of state $I$ in negative parity. $\Delta E(I)$ is the parity splitting~\cite{JolosR2011_PRC84_24312, JolosR2012_PRC86_24319},
\begin{equation}
   \Delta E(I) = E_{\rm exp}(I) - E_{\rm inter}(I),
\end{equation}

$E(\omega)= \langle \psi | H | \psi \rangle$, energy of state $|\psi\rangle$ in the PNC-CSM, is parity-independent function of the  rotational frequency $\omega$. Similar to Eq.~(\ref{eq:E_exp}), the positive- and negative-parity bands can be expressed as   
\begin{eqnarray}
E_p(\omega)=E(\omega)-\frac{1}{2}p\Delta E(\omega)\ .
\end{eqnarray}
where $\Delta E(\omega)$ is the parity splitting.  

The angular momentum alignment and moment of inertia for positive- and negative-parity bands have similar forms,
\begin{eqnarray}
    \langle J_x \rangle_{p} = \langle \psi | J_x | \psi \rangle - \frac{1}{2} p \Delta I_x(\omega),
     \label{eq:Jx_p}\\
    J^{(1)}_p= \frac{\langle \psi | J_x | \psi \rangle}{\omega} - \frac{1}{2} p \Delta J^{(1)}(\omega)
    \label{eq:J1_p}
\end{eqnarray}
$\Delta I_x(\omega)$ and $\Delta J^{(1)}(\omega)$ are parity splitting of the alignment and MoI, respective, which can be obtained from the experimental data as, 
\begin{eqnarray}
    \Delta I_x(\omega) = {I_x}_{-}(\omega)-{I_x}_{+}(\omega)\ ,
    \label{eq:delta_Jx}\\
    \Delta J^{(1)}(\omega) = J^{(1)}_{-}(\omega)-J^{(1)}_{+}(\omega).
    \label{eq:delta_J1}
  \end{eqnarray}
with $+(-)$ denoting the corresponding value of positive- (negative-) parity bands.    
  
\begin{figure}
    \includegraphics[scale=0.3]{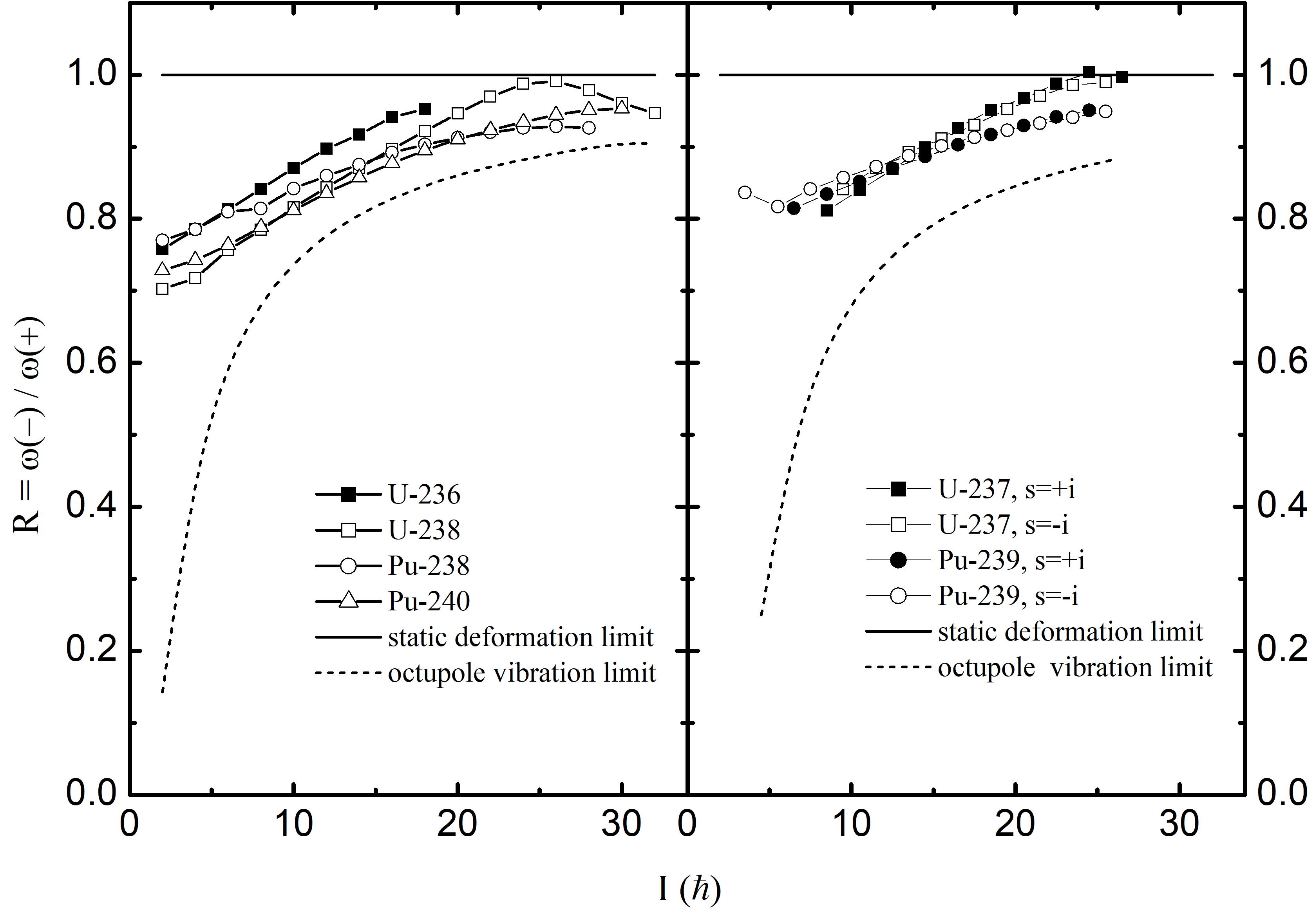}
    \caption{Ratio of rotational frequencies of the positive- and negative-parity bands $R=\omega(-)/\omega(+)$ as a function of angular momentum $I$ for even-even nuclei \element{236,238}{U} and \element{238,240}{Pu} (left) and odd-A nuclei \element{237}{U} and \element{239}{Pu} (right). The experimental data are taken from Refs.~\cite{BrowneE2006_NDS107_2649, BrowneE2015_NDS127_191,SinghB2008_NDS109_2439,ZhuS2010_PRC81_41306,WardD1996_NPA600_88}. The solid and dashed lines show the static octupole deformation limit and octupole vibration limit, respectively.}\label{fig:Fig1}
\end{figure}

\section{Experimental information of the yrast bands in U and Pu isotopes}
\label{sec:Exp}

Empirically, it shows that rotation can stabilize ocutpole deformation, namely octupole shapes are more stable at high spin than at low spin~\cite{AhmadI1993_ARNPS43_71, ButlerP1996_RMP68_349}. In a rotational band with same simplex, the ratio of rotational frequency of the negative parity band and the positive parity band is defined as~\cite{AhmadI1993_ARNPS43_71},
\begin{equation}
    R = \omega(-) / \omega(+).
\end{equation}

In the limit of static octupole deformation parity splitting should be vanished and at the meantime the ratio between the rotational frequency of the positive- and negative-parity bands should be close to one, $R\to R_{\rm{rigid}}=1$. Another limit is the limit of aligned octupole phonon, it is $R\to(4(I-3)-2)/(4I-2)$. Fig.\ref{fig:Fig1} plots the ratio $R$ versus $I$ for nuclei \element{236,237,238}{U} and \element{238,239,240}{Pu}. It shows that these nuclei have fine deformation stability and the octupole shapes are stabilized by rotation. $R$ approaches to $R=1$ at high spin for both of even-even nuclei \element{236,238}{U}, \element{238,240}{Pu} and odd-A nuclei \element{237}{U}, \element{239}{Pu}. 

More detailed, value of $R$ is bigger (closer to one) for U isotopes than that for Pu isotopes at the high spin region. This implies that there might be ocupole vibration mixed for the bands in Pu isotopes even at the high spin region. When compared values of $R$ at the low spin region, it found that $R_{\textrm{odd-A}}>R_{\textrm{even-even}}$. This means octupole deformation becomes more stable in odd mass nuclei due to the existent of the unpaired nucleon~\cite{HeX2006_IJMPE15_1823}. This issue needs further investigations. 

\section{Results and discussions}{\label{sec:results}}
\subsection{parameters}

The set of Nilsson parameters ($\kappa, \mu$) is taken from Ref.~\cite{ZhangZ2012_PRC85_14324}. The deformation parameters $\varepsilon_2$, $\varepsilon_3$ and $\varepsilon_4$ used in the calculations are listed in Table~\ref{tab:deformation}. The deformation parameters $\varepsilon_2$ and $\varepsilon_4$ are very close to the calculated ground-state deformations in actinide region~\cite{NilssonS1969_NPA131_1}, where $\varepsilon_2$ are little larger than those predicted in the macroscopic-microscopic models~\cite{SobiczewskiA2001_PRC63_343061} and the finite range droplet model~\cite{MollerP1995_ADNDT59_185}. The octupole deformation parameters $\varepsilon_3$ are chosen by fitting the experimental yrast bands of \element{236,238}{U} and \element{238,240}{Pu}.

\begin{table}
\caption{\label{tab:deformation} Deformation parameters $\varepsilon_2$, $\varepsilon_3$ and $\varepsilon_4$ used in the present PNC-CSM calculations for \element{236,238}{U} and \element{238,240}{Pu}. }
\begin{center}
\def\temptablewidth{0.47\textwidth}
\begin{ruledtabular}
\begin{tabular*}{\temptablewidth}{@{\extracolsep{\fill}}ccccccc} 
  & \element{236}{U}  & \element{238}{U} & \element{238}{Pu} & \element{240}{Pu} \\ \hline
  $\varepsilon_2$ & 0.200 & 0.220 & 0.228  & 0.230 \\
  $\varepsilon_3$ & 0.110 &  0.130 & 0.025  & 0.010 \\
  $\varepsilon_4$ & -0.055& -0.040 & -0.065 & -0.045 \\
\end{tabular*}
\end{ruledtabular}
\end{center}
\end{table}

The effective pairing strengths $G_0$ and $G_2$ can be determined by the odd-even differences in nuclear binding energies in principle. They also depends on the dimensions of the truncated CMPC space. In the present calculation, the CMPC space is constructed in the proton $N=5, 6$ and neutron $N=6, 7$ major shells, and the dimensions are about 1000 for both protons and neutrons. The effective pairing strengths are $G_{0p}=0.25{\rm MeV}$, $G_{2p}=0.03{\rm MeV}$ and $G_{0n}=0.25{\rm MeV}$, $G_{2n}=0.015{\rm MeV}$ for protons and neutrons, respectively. For the yrast and low-lying excited states, the number of important CMPC (weight $>10^{-2}$) is very limited ($<20$), and almost all of CMPC with weight $>10^{-3}$ are taken into account. The PNC-CSM calculations are stable against the change of the dimension of the CMPC space, and calculations in a larger CMPC space with decreased effective pairing strengths give the result nearly unchanged~\cite{LiuS2002_PRC66_24320}.

\subsection{Cranked Nilsson levels}

\begin{figure*}[ht]
    \includegraphics[scale=0.35]{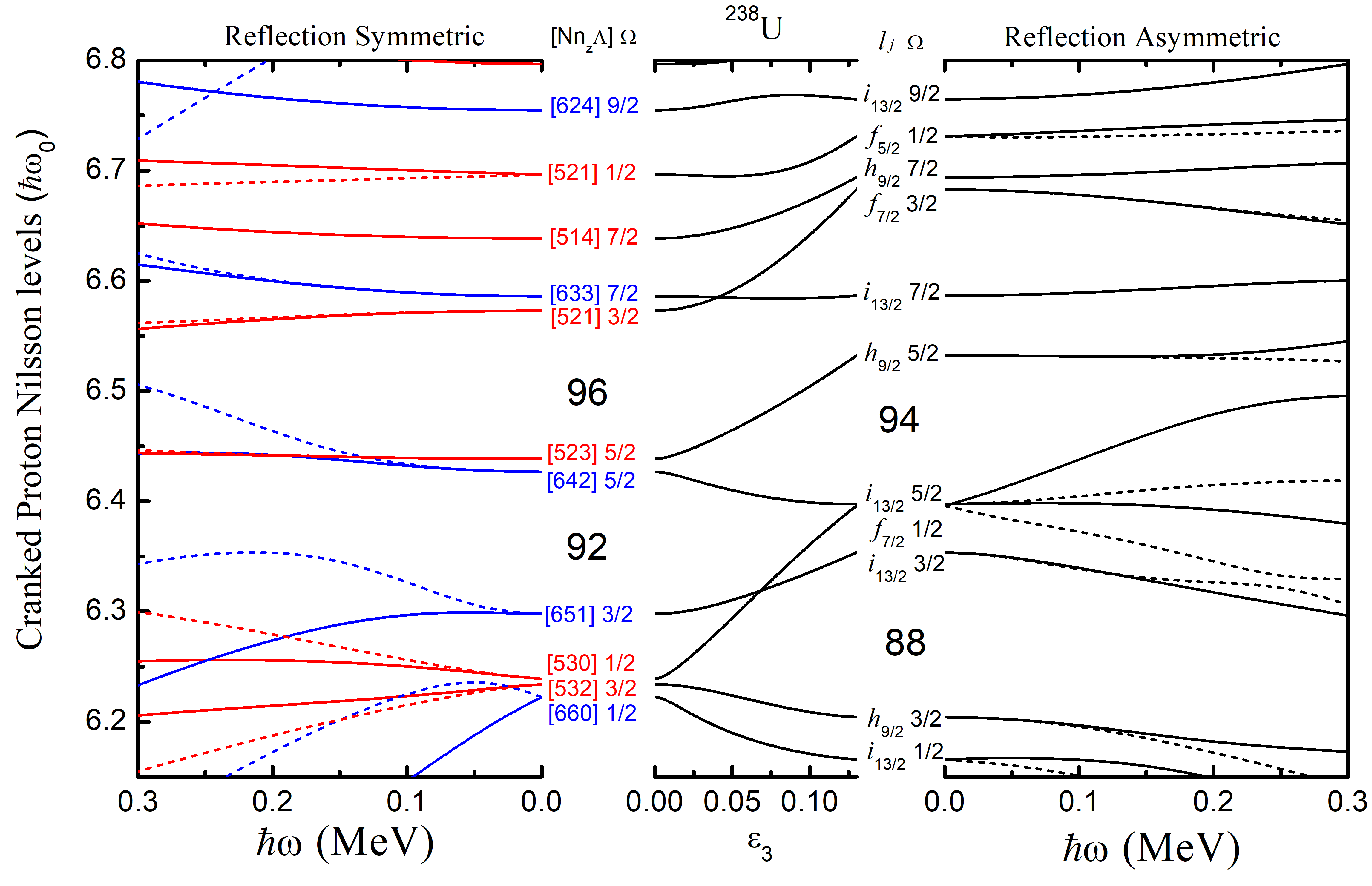}
    \includegraphics[scale=0.35]{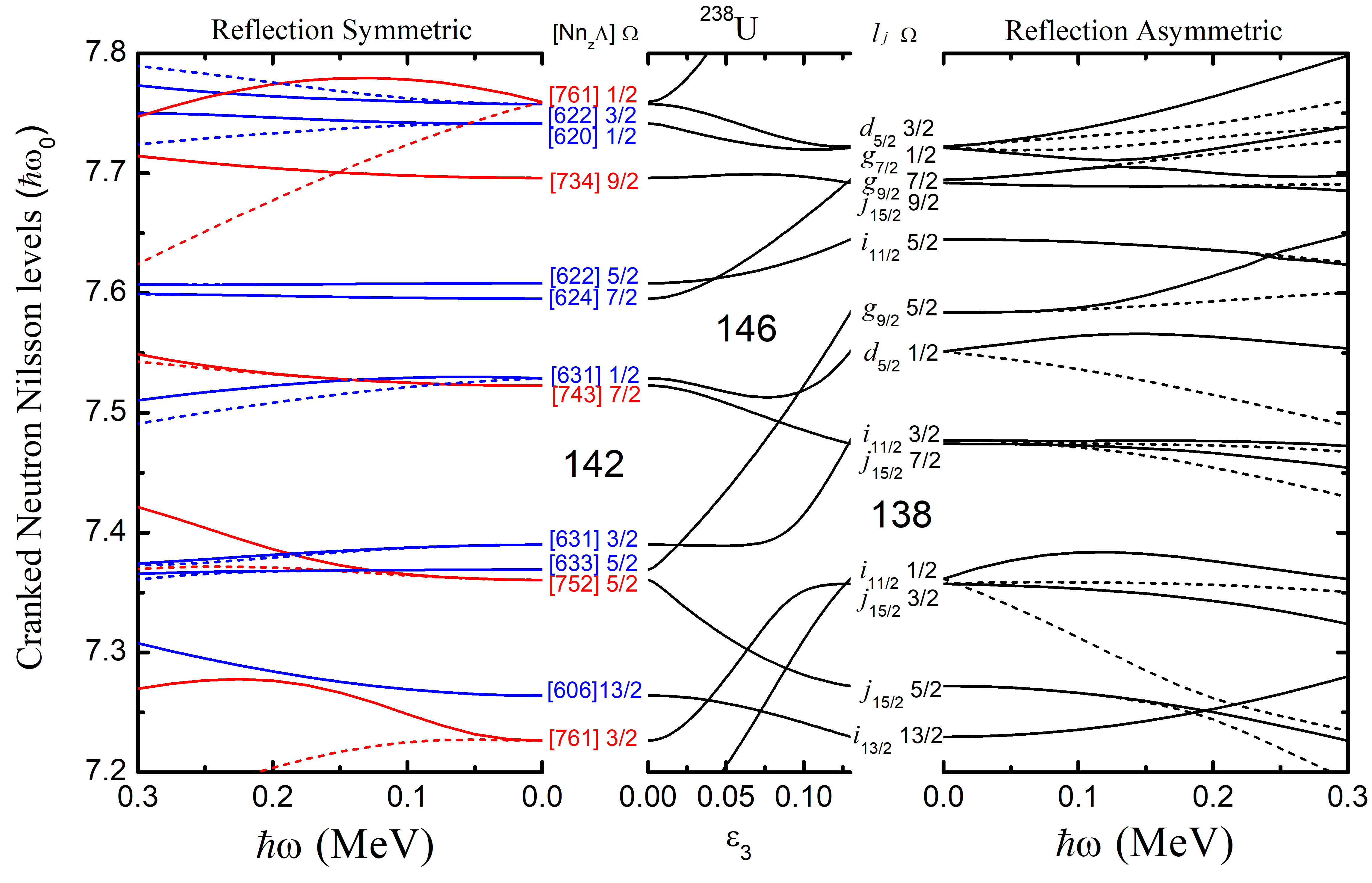}
    \caption{\label{fig:Fig2}
    Cranked Nilsson levels near the Fermi surface of \element{238}{U} for proton (top) and neutron (bottom). For reflection symemetric system ($\varepsilon_3=0$), the positive (negative) parity levels are denoted by blue (red) lines with quantum numbers $[Nn_z\Lambda]\Omega$ and $\alpha=+1/2$ ($\alpha=-1/2$) signatures levels are denoted by solid (dashed) lines. For the reflection asymmetric system ($\varepsilon_3\neq0$), the simplex $s=+i$ ($s=-i$) levels are denoted by black solid (dashed) lines with quantum numbers $l_j\Omega$.}
\end{figure*}

\begin{figure*}[ht]
    \includegraphics[scale=0.35]{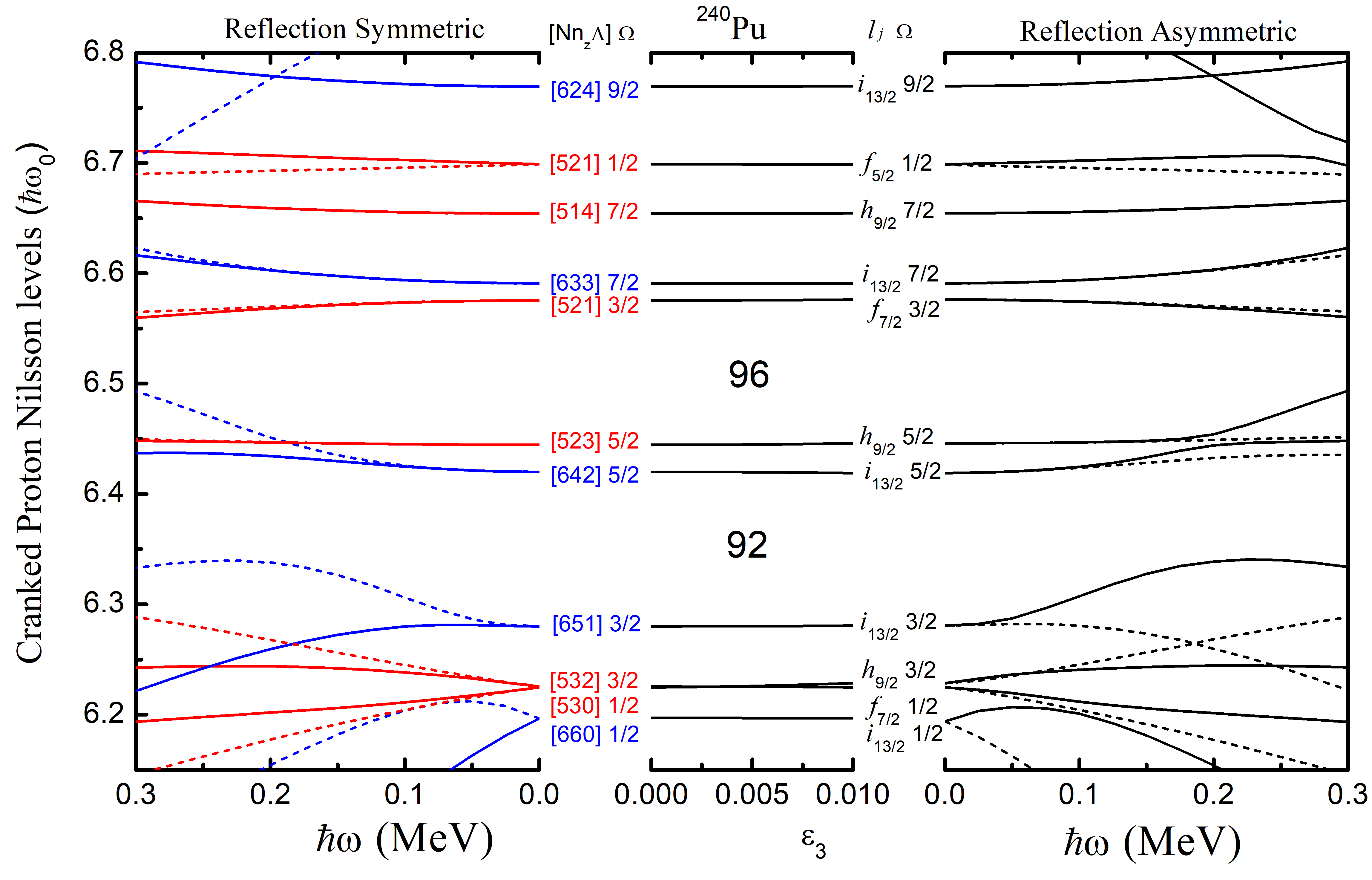}
    \includegraphics[scale=0.35]{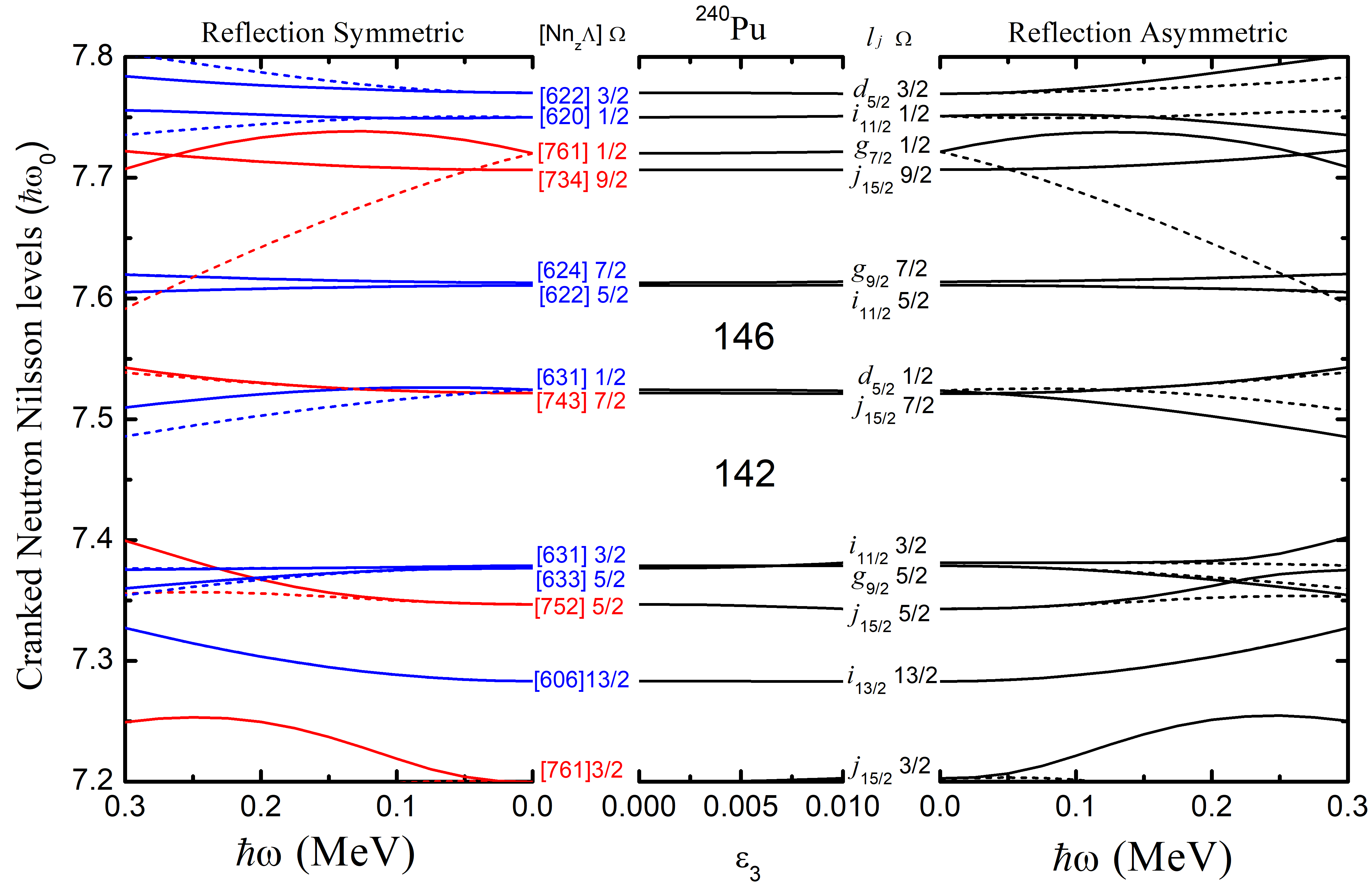}
    \caption{\label{fig:Fig3}
    The same as Fig.\ref{fig:Fig2}, but for cranked Nilsson levels near the Fermi surface of \element{240}{Pu}.}
\end{figure*}

Figures~\ref{fig:Fig2} and~\ref{fig:Fig3} show the calculated cranked Nilsson levels near the Fermi surface of \element{238}{U} and \element{240}{Pu}, respectively. When $\varepsilon_3=0$,
the positive (negative) parity levels are denoted by blue (red) lines, and the signature $\alpha=+1/2$  ($-1/2$) levels are denoted by solid (dotted) lines with quantum numbers $[Nn_z\Lambda]\Omega$ at the band head ($\omega=0$). When $\varepsilon_3\neq0$, the levels are denoted by black lines, and the simplex  $s=+i$ ($-i$) levels are denoted by solid  (dotted) lines with quantum number $\Omega$ at the band head ($\omega=0$). The cranked Nilsson levels near the Fermi surface of \element{236}{U} and \element{238}{Pu} are quite similar to that of \element{238}{U} and \element{240}{Pu}, respectively, and will not be displayed here. 

Based on such a sequence of single-particle levels, the experimental ground state and low-lying excited states in their neighbor odd-A nuclei can be reproduced well, such as the proton exited states in \element{237}{Np} and \element{241}{Am}~\cite{AbuSaleemK2004_PRC70_24310} and neutron exited states in \element{237}{U} and \element{239}{Pu} except for the first exited state with configuration of $\nu[622]5/2^+$. The disagreement of the position of $\nu[622]5/2^+$ in the theoretical prediction and experimental data were discussed in Refs.~\cite{ZhangZ2012_PRC85_14324, ZhangZ2011_PRC83_11304R}. The $Z=92,96$ gaps for protons and the $N=142,146$ gaps for neutrons in the reflection-symmetric deformed field is consistent with the calculation by using a Woods-Saxon potential~\cite{ChasmanR1977_RMP49_833}. The $Z=88,94$ gaps for protons and $N=138,142$ gaps for neutrons in the octupole deformed field is consistent with results of a Woods-Saxon potential~\cite{LeanderG1988_PRC37_2744} and a folded Yukawa potential~\cite{JainA1990_RMP62_393}.

Comparing Fig.\ref{fig:Fig2} with Fig.\ref{fig:Fig3}, it is seen that the cranked Nilsson levels in a stronger octupole deformed field of \element{238}{U} are quite different from that of \element{240}{Pu}. For \element{240}{Pu}, the cranked single-particle levels in reflection-asymmetric deformed field are similar to that in reflection-symmetric deformed field. As for \element{238}{U}, the proton $\pi 1/2$ level stemming from $ f_{7/2}$ orbital and the neutron $\nu 5/2$ level from $g_{9/2}$ orbital rise quickly to the Fermi surface as octupole deformation increases. As we know, the octupole correlation in this mass region is mainly concerning about nucleon occupying the octupole-correlation pairs of neutron $\nu^2 j_{15/2} g_{9/2}$ and of proton $\pi^2 i_{13/2} f_{7/2}$. Therefore, the properties of the rotational bands are influenced intensively by the octupole correlations in \element{236,238}{U}. 

\subsection{\label{sec:even-even}Alternating-parity bands in \element{236,238}{U} and \element{238,240}{Pu}}

\begin{figure*}
\includegraphics[scale=0.38]{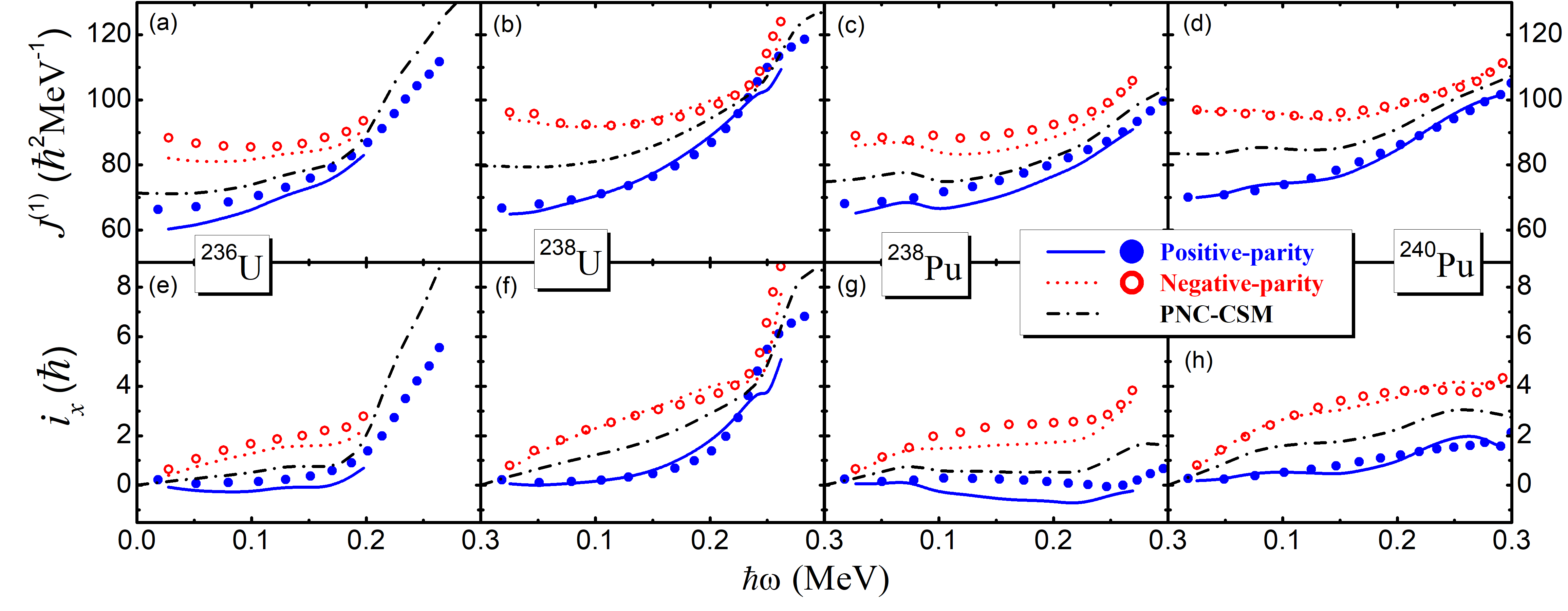}
\caption{\label{fig:Fig4}
The kinematic moments of inertia $J^{(1)}$ (top row) and alignments $J_x$ (bottom row) of the alternating-parity rotational bands in \element{236,238}{U} and \element{238,240}{Pu}. The experimental data are denoted by solid and open circles for the positive- and negative-parity bands, respectively, which are taken from Refs.~\cite{BrowneE2006_NDS107_2649, BrowneE2015_NDS127_191,SinghB2008_NDS109_2439,ZhuS2010_PRC81_41306,WardD1996_NPA600_88}. The alignment $i_x$ are obtained by subtracting a common reference $ \omega J_0 +\omega^3 J_1$ where Harris parameters $J_0=65\hbar^2{\rm MeV}^{-1}$ and $J_1=369\hbar^4{\rm MeV}^{-3}$ are taken from Ref.~\cite{ZhuS2005_PLB618_51}. The PNC-CSM calculations of the ground-state bands are denoted by dashed-dotted lines. Considering the parity splitting, the positive- and negative-parity bands are denoted by solid and dotted lines, respectively.}
\end{figure*}

The kinematic moments of inertia and alignments of the ground-state bands of \element{236,238}{U} and \element{238,240}{Pu} are shown in Fig.\ref{fig:Fig4}, which show an alternating-parity structure. The PNC-CSM calculations of $J_x$ [Eq.~(\ref{eq:jx})] and $J^{(1)}$ [Eq.~(\ref{eq:moi})] are presented by dashed-dotted lines. Considering the parity splitting [Eq.~(\ref{eq:delta_Jx}) and Eq.~(\ref{eq:delta_J1})], the alternating-parity bands are shown as the solid and dotted lines for positive- and negative-parity bands, respectively. Experimental data are denoted by solid and open circles for positive- and negative-parity bands, respectively, which are taken from Refs.~\cite{BrowneE2006_NDS107_2649, BrowneE2015_NDS127_191, SinghB2008_NDS109_2439,ZhuS2010_PRC81_41306}. The experimental MoIs and alignments are reproduced very well by the PNC-CSM calculation. It is seen that the rotational behaviors of U isotopes and Pu isotopes are quite different. There are distinct upbendings for both alternating-parity bands in \element{236,238}{U} while its behaviors are much plain for bands in \element{238,240}{Pu}. It is known that the backbending is caused by crossing of the ground-state band with a pair-broken band based on the high-$j$ intruder orbitals. In this region, the high-$j$ intruder orbitals near the Fermi surface are the proton $\pi  i_{13/2}$ and neutron $\nu  j_{15/2}$ orbitals. 

\begin{figure*}
\includegraphics[scale=0.55]{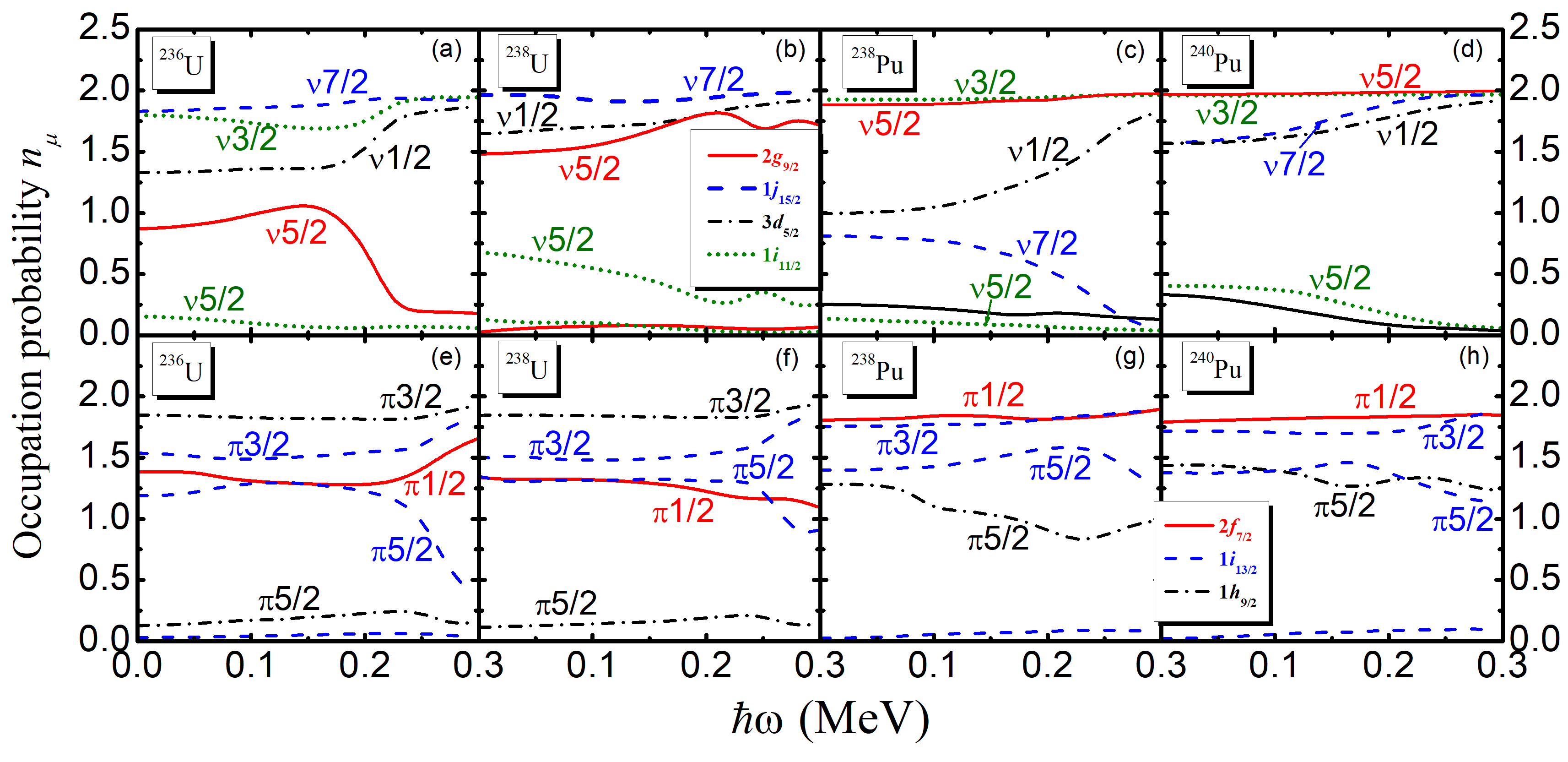}
\caption{\label{fig:Fig5}
    The occupation probability $n_\mu$ of each neutron (top row) and proton (bottom row) orbitals $\mu$  (including both $s=\pm i$) near the Fermi surface for the alternating-parity bands in \element{236,238}{U} and \element{238,240}{Pu}. The Nilsson levels far above ($n_\mu\approx0$) and far below ($n_\mu\approx2$) the Fermi surface are not shown.}
\end{figure*}

Figure~\ref{fig:Fig5} shows the occupation probability $n_\mu$ of each orbital $\mu$  (including both $s=\pm i$) near the Fermi surface for the alternating-parity bands in \element{236,238}{U} and \element{238,240}{Pu}. For U isotopes, due to the effect of octupole correlation, proton $\pi1/2$ and neutron $\nu5/2$ orbitals rise to the Fermi surface rapidly (see Fig.~\ref{fig:Fig2}). As shown in Fig.~\ref{fig:Fig5}, both $\pi1/2$ and $\nu5/2$ are partially occupied. The octupole correlation between pairs of nucleons occupying  $\nu^2 j_{15/2} g_{9/2}$ and of $\pi^2 i_{13/2} f_{7/2}$ orbitals will affect strongly the rotational properties of the alternating-parity bands in U isotopes.  

\begin{figure*}
\includegraphics[scale=0.55]{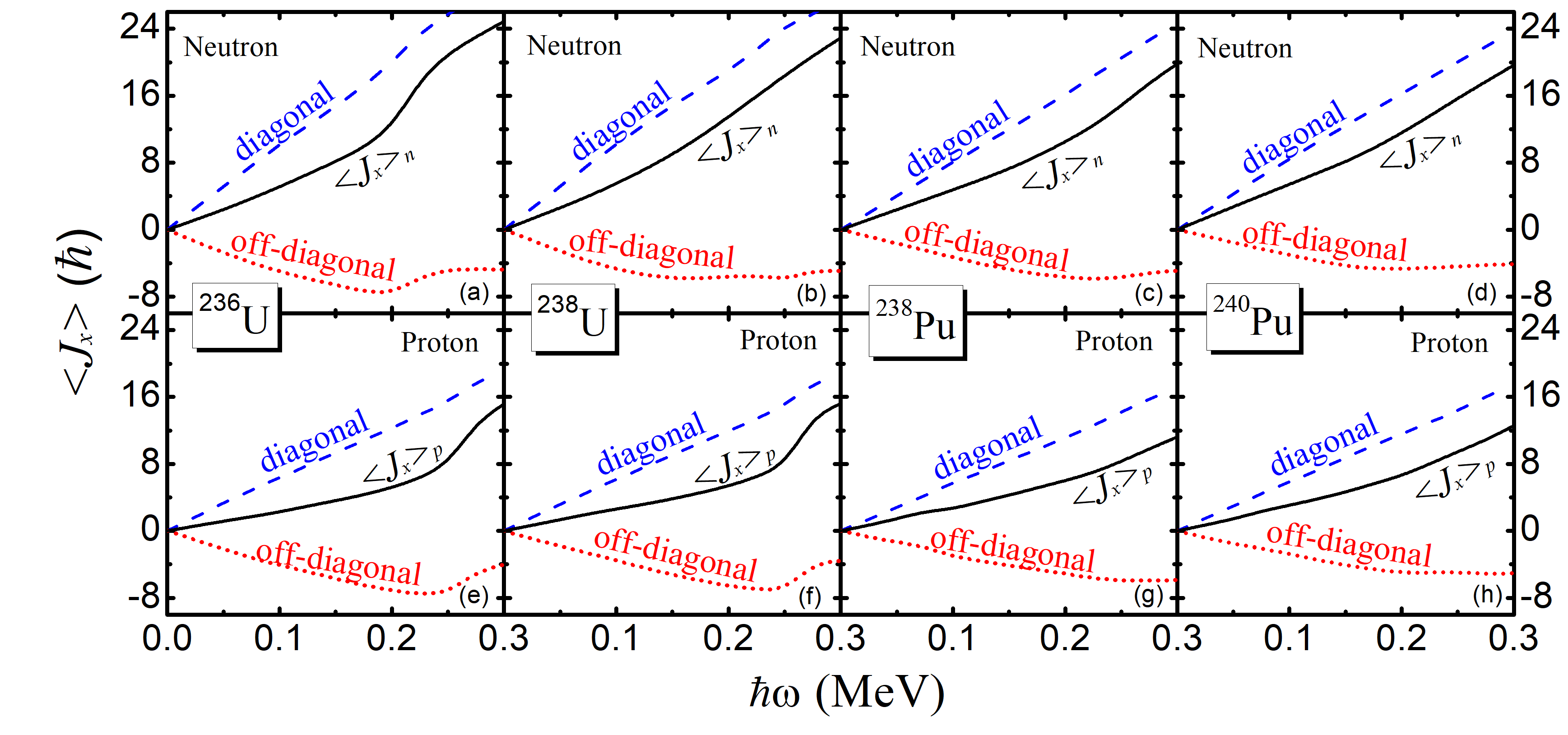}
\caption{\label{fig:Fig6}
    Contributions of neutron (top row) and proton (bottom row) to the angular momentum alignment $\langle J_x\rangle$ for the alternating-parity rotational bands in \element{236,238}{U} and \element{238,240}{Pu}. The diagonal $\sum_{\mu} j_x (\mu)$ and off-diagonal parts $\sum_{\mu<\nu} j_x (\mu\nu)$ are denoted by blue dashed and red dotted lines, respectively.}
\end{figure*}

Upbendings of the alternating-parity bands occur at frequency $\hbar\omega\approx0.20$ MeV in \element{236}{U} while that are delayed to $\hbar\omega\approx0.25$ MeV in \element{238}{U}. From Fig.~\ref{fig:Fig5} (e) and (f), we can see that the proton occupation probability for \element{236}{U} and \element{238}{U} are very similar. Partially occupation of orbitals $\pi5/2$, $\pi1/2$ and $\pi3/2$ are almost constant at $\hbar\omega<0.25$ and changed rapidly at $\hbar\omega>0.25$. This leads to upbendings of proton alignment at $\hbar\omega\approx0.25$ [see Fig.~\ref{fig:Fig6} (e) and (f)]. Occupation probabilities of neutrons for \element{236}{U} and \element{238}{U} are quite different. As shown in Fig.~\ref{fig:Fig5} (a), both of $\nu1/2$ and $\nu5/2$ are half occupied ($n_{\mu}\approx1$) at $\hbar\omega<0.20$ MeV and $\nu1/2$ gets nearly fully occupied ($n_{\mu}\approx2$) while $\nu5/2$ becomes almost empty ($n_{\mu}\approx0$) at $\hbar\omega>0.20$ MeV for \element{236}{U}. Meanwhile, both of $\nu1/2$ and $\nu5/2$ are almost fully occupied and keep nearly constant with $n_{\mu}=1.5-2.0$ at the whole frequency region for \element{238}{U} [see Fig.~\ref{fig:Fig5} (b)]. Therefore, as shown in Fig.~\ref{fig:Fig6} that upbendings of alternating-parity bands in \element{236}{U} are mainly due to the sudden increased neutron alignment at $\hbar\omega\approx0.20$ MeV while ones in \element{238}{U} are mostly from the rapidly gained proton alignment at $\hbar\omega\approx0.25$ MeV. This difference is easy to understand since the neutron Fermi surface of \element{238}{U} locates just above the $N=146$ deformed shell. 

For both of \element{238,240}{Pu}, the moments of inertia $J^{(1)}$ keep nearly constant at frequency $\hbar\omega<0.2$ MeV and increased slightly at $\hbar\omega>0.2$ MeV. Since neutron $\nu5/2$ ($ g_{9/2}$) and proton $\pi1/2$ ($ f_{7/2}$) levels locate well below the Fermi surface, both of neutron $\nu5/2$ and proton $\pi1/2$ orbitals are nearly fully occupied with $n_{\mu}\approx2$, while the high-$j$ orbitals $\nu7/2$ $( j_{15/2})$ and $\pi5/2$ $( i_{13/2})$ are partially occupied. Thus, unlike \element{236,238}{U}, in which upbendings of $J^{(1)}$ are effected strongly by the octupole correlation between $\nu^2 j_{15/2} g_{9/2}$ pairs and proton $\pi^2 i_{13/2} f_{7/2}$ pairs, in \element{238,240}{Pu}, it is the high-$j$ intruder orbitals $\nu j_{15/2}$ and $\pi i_{13/2}$ that influences the variation of $J^{(1)}$ versus frequency. 

\begin{figure*}
\includegraphics[scale=0.55]{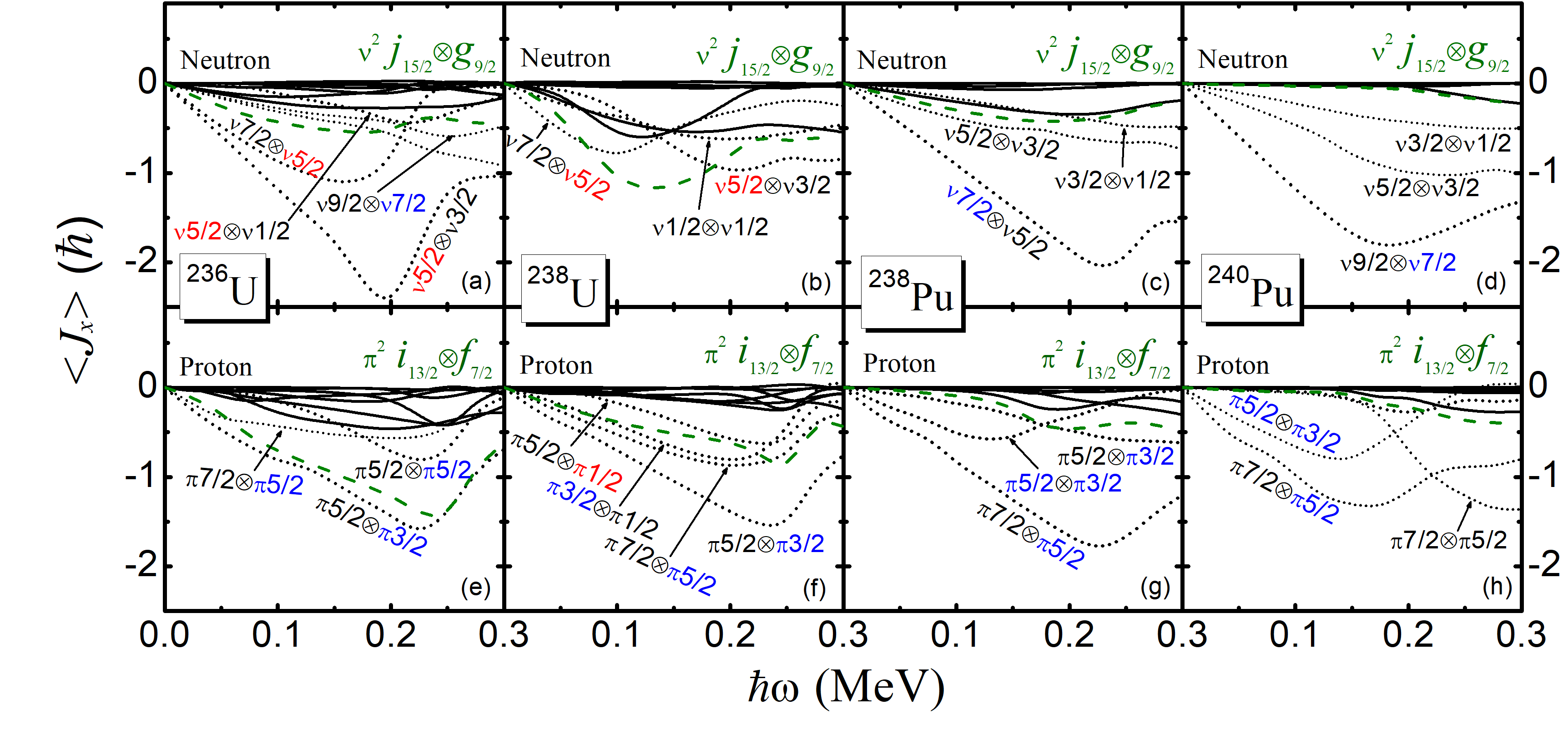}
\caption{\label{fig:Fig7}
    The off-diagonal parts $j_x (\mu\nu)$ of contribution from each neutron (top row) and proton (bottom row) orbitals to the angular momentum alignments $\langle J_x\rangle$ for the alternating-parity rotational bands in \element{236,238}{U} and \element{238,240}{Pu}.
     The interference term $j_x (\mu\nu)$ between orbitals from neutron $\nu^2  j_{15/2}  g_{9/2}$ paris and proton $\pi^2  i_{13/2}  f_{7/2}$ paris are denoted by black solid lines, sum of which are denoted by olive dashed lines. Other interference terms are denoted by black dotted lines. The orbitals that have little contributions are not shown. Orbitals from $\nu g_{9/2}$ ($\pi f_{7/2}$) and $\nu  j_{15/2}$ ($\pi  i_{13/2} $) are denoted by red and blue quantum number $\Omega$, respectively.} 
\end{figure*}

The contributions of proton (bottom row) and neutron (top row) to the angular momentum alignment $\langle J_x\rangle$ for the alternating bands in \element{236,238}{U} and \element{238,240}{Pu} are shown in Fig.\ref{fig:Fig6}. The diagonal $\sum_{\mu} j_x (\mu)$ and off-diagonal $\sum_{\mu<\nu} j_x (\mu\nu)$ parts are denoted by blue dashed and red dotted lines, respectively. In general, the gradual rise of $J^{(1)}$ for all the studied bands are attributed to diagonal parts of both neutron and proton alignment. The upbending of $J^{(1)}$ at $\hbar\omega\approx0.20$~MeV in \element{236}{U} is mainly due to the off-diagonal contribution from neutrons while one at $\hbar\omega\approx0.25$~MeV in \element{238}{U} is from the off-diagonal part of proton alignment. As for \element{238,240}{Pu}, only very subtle increases (stops falling actually) happen for off-diagonal parts of both neutrons and protons alignments, which result to the slight increases of $J^{(1)}$ at $\hbar\omega>0.20$ MeV. 

The contributions to the alignment from each single-particle levels are shown in Fig.~\ref{fig:Fig7}. According to Eq.~(\ref{eq:Jx_single}), it includes the direct term $j_x({\mu})$ and the interference term $j_x (\mu\nu)$. As discussed above, only the off-diagonal parts contribute to upbendings of moment of inertia versus frequency. Therefore, only the interference terms are displayed. 

In Fig.~\ref{fig:Fig7}, the interference term $j_x (\mu\nu)$ between orbitals from neutron $\nu^2  j_{15/2}  g_{9/2}$ paris and proton $\pi^2  i_{13/2}  f_{7/2}$ paris are denoted by black solid lines. Other interference terms, which have importance effect on the alignment, are denoted by black dotted lines. The orbitals that have little contributions are not shown. To investigate the impact of the octupole correlation on the rotational properties, all terms of $j_x (\mu\nu)$ (black solid lines) belonging to the neutron $\nu^2  j_{15/2} g_{9/2}$ paris and proton $\pi^2  i_{13/2} f_{7/2}$ paris are added, sum of which are denoted by olive dashed lines. 

It can be seen clearly that the upbending of the alternating-parity bands at $\hbar\omega\approx0.20$ MeV in \element{236}{U} mostly attribute to the alignments of neutrons occupying orbital $\nu5/2$ $( g_{9/2})$ and high-$j$ intruder orbital $\nu7/2$ $( j_{15/2})$. Particularly, the interference terms between neutron $\nu^2  j_{15/2}g_{9/2}$ paris give a considerable contribution to the suddenly increased alignment at $\hbar\omega\approx0.20$ MeV. 

For \element{238}{U}, the upbendings of $J^{(1)}$ are mainly due to the suddenly gained alignment of protons occupying otbital $\pi1/2 $ $( f_{7/2})$ and high-$j$ intruder orbitals $\pi3/2$ $( i_{13/2})$ and $\pi5/2$ $( i_{13/2})$. As shown in Fig.~\ref{fig:Fig7} (f) that the interference terms (olive dashed line) between $\pi^2  i_{13/2}  f_{7/2}$ paris play a very important role in the sharp increased alignment. For neutrons, although the alignment of $\nu^2  j_{15/2}  g_{9/2}$ rises suddenly at $\hbar\omega\approx0.15$ MeV, the effect is cancelled out by contributions from other orbitls. 

For \element{238,240}{Pu}, only the interference terms $j_x (\mu\nu)$ concerning the high-$j$ orbitals $\nu7/2$ $( j_{15/2})$ and $\pi5/2$ $( i_{13/2})$ increases a little at the high frequency region, and ones from neutron $\nu^2  j_{15/2}  g_{9/2}$ pairs and  $\pi^2  i_{13/2}  f_{7/2}$ paris give little contributions. Therefore, the moment of inertia of the alternating-parity bands in \element{238,240}{Pu} are nearly constant at frequency $\hbar\omega<0.2$ MeV and increased slightly at $\hbar\omega>0.2$ MeV. 

\section{Summary}{\label{sec:summary}}

The particle-number-conserving pairing method in the framework of cranked shell model is developed to treat the reflection-asymmetric nuclear system by including the octupole deformation. Based on an octupole-deformed Nilsson potential, the alternating-parity bands in even-even nuclei \element{236,238}{U} and \element{238,240}{Pu} have been studied. The observed $\omega$ variations of moment of inertia $J^{(1)}$ and the angular momentum alignments of all studied bands are reproduced very well by the PNC-CSM calculations. The significant difference of rotational properties between U and Pu isotopes are explained. 

For all the studied bands in the present work, it is the off-diagonal parts of the alignment effect mostly the variation of the moment of inertia $J^{(1)}$ versus frequency. The diagonal parts of alignment contribute mainly to the gradual rise of the moment of inertia $J^{(1)}$. 

The octupole correlation for U and Pu isotopes in this region is mainly concerning with nucleons occupying pairs of neutron $\nu^2 j_{15/2} g_{9/2}$ and proton $\pi^2 i_{13/2} f_{7/2}$ orbitals. The upbending of the alternating-parity bands in \element{236}{U} is mainly due to the suddenly gained alignment of nucleons occupying the neutron $\nu5/2 $ $(g_{9/2})$ and $\nu7/2$ $( j_{15/2})$ orbitals while one in \element{238}{U} attributes to sharp increased alignment of nucleon occupying the proton $\pi1/2$ $( f_{7/2})$ and $\pi3/2$ $ (i_{13/2})$ orbitals. Particularly, the interference terms of the alignment for nucleons occupying the octupole-deformed pairs of neutron $\nu^2 j_{15/2} g_{9/2}$ and of proton $\pi^2 i_{13/2} f_{7/2}$ orbitals give a very important contribution to the upbendings. 

Compared to the case of \element{236}{U}, the upbending frequency of $J^{(1)}$ is delayed to the higher frequency region in \element{238}{U}. This is because the neutron Fermi surface of \element{238}{U} locates just above the deformed sub-shell at $N=146$. Then the upbendings of $J^{(1)}$ is mostly from the contribution of proton alignments which happen at higher frequency.  

Variation of $J^{(1)}$ versus frequency is much plain for the alternating-parity bands in \element{238,240}{Pu}, which can be reproduced and explained based on a Nilsson potential with comparatively weaker octupole correlations. Under a weaker octupole-deformed field, $\nu5/2$ $ ( g_{9/2})$ and $\pi1/2$ $( f_{7/2})$ orbitals are well below the Fermi surface.  Alignments from nucleons occupying the octupole correlation pairs of $\nu^2 j_{15/2} g_{9/2}$ and of $\pi^2 i_{13/2} f_{7/2}$ are very trivial. Then the alignment are mainly gained from nucleons occupying the high-$j$ intruder orbitals $\nu7/2$ $( j_{15/2})$, $\pi3/2$ $ ( i_{13/2})$ and $\pi5/2$ $ ( i_{13/2})$ at the high spin region, and its changes are quite gentle. 
 
\begin{acknowledgements}
This work is supported by the National Natural Science Foundation of China under Grant No. 11775112.
\end{acknowledgements}


\bibliographystyle{apsrev4-1}
\bibliography{../../../References/ReferencesXT}

\end{document}